# Some Aspects of Dynamical Tunneling in Two Dimensional Coupled States

Dmytro Babyuk

*Department of Chemistry, University of Nevada, Reno, Reno Nevada 89557*

Process of dynamical tunneling in two-dimensional coupled potentials is considered within Bohmian approach to quantum mechanics. Quantum trajectories tend to go along the paths where potential energy increases and then decreases. It leads to a suggestion that dynamical and barrier tunneling are the same processes but the former is hidden due to specific potential surface.

## I. INTRODUCTION

Classical study for the vibrations of the water molecule by Lawton and Child has revealed an existence of local modes [1]. Classical trajectories may appear nonsymmetric even in symmetric potential, reflecting the local structure of vibrations. The locality means that the eigenstates are supposed to be split. Later the authors have examined it quantum mechanically and found that the eigenstates appear to be doublets and their splitting decreases with the increase of vibrational quantum number [2]. The superposition of low doublet splitting states create localized ones in bond directions. These local states can make transitions between each other and the splitting is a measure of the energy transfer rate. Such transition between local states is classically forbidden. The system behaves like being under the influence of the double well potential though there is no potential barrier. A similar result has been obtained by Davis and Heller in a study of quantum system in two-dimensional coupled harmonic potential [3]. They proposed to call that process dynamical tunneling because of the possibility of transition of quantum system between classically trapped regions of phase space in the absence of potential barrier. Dynamical tunneling plays an important role in many processes [4,5] and cannot be ignored in the semiclassical method. Moreover, it is harder to reveal dynamical tunneling than barrier tunneling because the barrier is not obvious from the potential surface. Instead, classical trajectories must be examined.

From semiclassical point of view the barrier tunneling in double well potential causes the energy splitting [6]. It was also noted in previous works about dynamical tunneling. In current work we try to clarify why the energy splitting is available in the absence of barrier in dynamical tunneling process.

As a matter of fact, it is not necessary for quantum system to go along the path where potential energy is lower or equal to initial one, even if there is no potential restriction. An assumption was made that during dynamical tunneling the probability density may flow along the path where potential energy increases and then decreases. Therefore this fact can be treated as barrier tunneling.

Only an analysis of wave function evolution as a whole is incapable of demonstration of barrier tunneling in this case. The problem should be solved by decomposition of initial density distribution into components and examination of their motion. From that result one can conclude if the barrier tunneling occurs. The approach described above is a causal interpretation of quantum mechanics introduced by Bohm [7-9].

The remainder of the paper is organized as follows. In Sec.II, we briefly present the basic idea of Bohmian approach for two-dimensional case. Then we apply it for dynamical tunneling for coupled harmonic (Sec.III) and water (Sec.IV) potential. Finally, we summarize our findings in Sec.V.

## II. BOHMIAN APPROACH

The Bohmian approach is an alternative interpretation of quantum mechanics. According to it a wave propagation can be represented as a motion of ensemble of interacting point particles. Each particle obeys the law of motion which is derived from the Schroedinger equation and follows a definite track in space and time. Therefore this approach is also called causal interpretation of quantum mechanics. A motion of all particles reveals a wave function evolution. A probability conception is not so important here as in usual interpretation but it can still be defined as a value proportional to particle density in space.

As noted above, such an approach is useful for us because it explores the motion of a single component of wave function and then one can extract the information if some pieces of wave packet tend to barrier tunneling.

Mathematically, the main idea of Bohmian approach is based on representation of wave function in a polar form (two dimensions) [7-9]

$$\Psi(x,y,t) = R(x,y,t)e^{iS(x,y,t)} \qquad (1)$$

where *R* and *S* are real values. Substitution of (1) in the time dependent Schroedinger equation with further separation of real and imaginary parts gives the system

$$-\frac{\partial S}{\partial t} = \frac{1}{2}\left(\frac{\partial S}{\partial x}\right)^2 + \frac{1}{2}\left(\frac{\partial S}{\partial y}\right)^2 + U(x,y) + Q(x,y,t) \quad (2)$$

$$\frac{\partial R^2}{\partial t} = -\frac{\partial}{\partial x}\left(R^2 \frac{\partial S}{\partial x}\right) - \frac{\partial}{\partial y}\left(R^2 \frac{\partial S}{\partial y}\right) \quad (3)$$

Equation (2) reminds a classical Hamilton-Jacobi equation with an additional term $Q$ called quantum potential. The phase $S$ is an analogue of classical action. (3) is a probability conservation equation. This system is equivalent to the Schroedinger and does not provide any additional information yet.

Considering (3) as

$$\frac{\partial R^2}{\partial t} = -\frac{\partial}{\partial x}(J_x) - \frac{\partial}{\partial y}(J_y)$$

where $J_x$ and $J_y$ – are fluxes and keeping in mind that they are a product of density and local velocity

$$J_x = R^2 v_x \quad J_y = R^2 v_y$$

the following equations are derived

$$v_x = \frac{\partial S}{\partial x}, \qquad v_y = \frac{\partial S}{\partial y}$$

Now substituting velocity for position derivative we obtain the equations for point particle motion

$$\begin{aligned}\frac{dx}{dt} &= \frac{\partial S}{\partial x} \\ \frac{dy}{dt} &= \frac{\partial S}{\partial y}\end{aligned} \quad (4)$$

If the initial condition for a particle is specified, then integration of the last system gives the trajectory in space which a particle explores.

Two difficulties may arise in the integration process. Firstly, if the time dependent wave function is unknown, then the phase $S$ has to be derived from eq. (2)-(3). In general, it is not easy to perform numerical integration due to quantum potential calculation. But if $\Psi$ is known, then $S$ is derived as

$$S = \arg(\Psi) \quad (5)$$

Secondly, the low energy splitting means that initial local state evolves too long and numerical integration of (4) is not possible to carry out for such long period of time, especially, if the trajectory encounters points near to nodes.

To avoid the first problem, it is desirable to have definite time dependent wave function. Another problem can be eliminated by performing integration only for relatively short time intervals. As a matter of fact, it is not necessary to have trajectory for the whole period of tunneling. Knowledge of potential change only for short time along trajectory may be enough.

In the next two sections we will apply the described approach to systems for which dynamical tunneling has been detected.

### III. COUPLED HARMONIC POTENTIAL

Davis and Heller demonstrated the existence of dynamical tunneling using the following potential [3]

$$U(x,y) = \frac{1}{2}\omega_x^2 x^2 + \frac{1}{2}\omega_y^2 y^2 + \lambda x y^2 \quad (6)$$

where parameters used for calculation were $\omega_x=1$, $\omega_y=1.1$, and $\lambda=-0.11$. Diagonalization of the potential (6) gives pairs of symmetric and antisymmetric states whose splitting tends to decrease if the energy increases. The sum and difference of split states create symmetric local states. One of those pairs with vibrational quantum numbers $v=15$ and $v=16$ is shown in Fig.1.

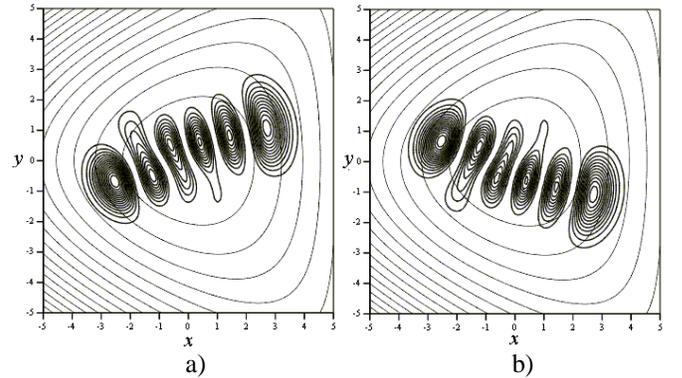

**Figure 1.** Local states composed of the sum (a) and difference (b) of split states $\psi_{15}$ ($E_{15}$=6.026) and $\psi_{16}$ ($E_{16}$=6.087).

These local states are not stationary and if one of them is left on its own, it evolves in time and finally transforms into its symmetric state. Thus dynamical tunneling happens. The time evolution is described by a simple relation

$$\Psi(x,y,t) = \frac{1}{\sqrt{2}}\left(\psi_s(x,y)e^{-iE_s t} + \psi_a(x,y)e^{-iE_a t}\right) \quad (7)$$

The motion is periodic with period

$$\tau = \frac{E_a - E_s}{2\pi} \quad (8)$$

At this stage we can apply the Bohmian approach for a study of dynamical tunneling of this system. Having the time dependent wave function (7), its phase is recovered according to (5) and specifying the initial condition out of node, the numerical integration of system (4) can be carried out. Fortunately, at our chosen parameters the tunneling period is $\tau = 103.58$ and that allows us to integrate the system (4) through the whole period. Using forth-order Runge-Kutta method for integration of (4), quantum trajectories were obtain. Some of them are depicted in Fig.2.

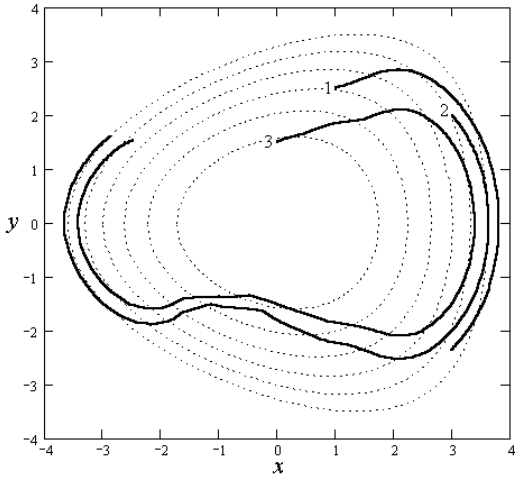

**Figure 2.** Trajectories of first type starting from the initial points: 1) $x(0)=1$, $y(0)=2.5$; 2) $x(0)=3$, $y(0)=2$; 3) $x(0)=0$, $y(0)=1.5$

For half-period of tunneling a trajectory makes an open curve and then returns to the starting point during the next half-period of tunneling. Such type of trajectory is not unique. Some trajectories make a closed line and circulate few times for half-period of tunneling. They are depicted in Fig.4.

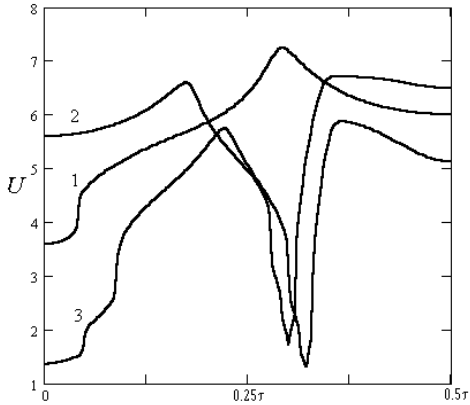

**Figure 3.** Potential energy evolution along the trajectories depicted in Fig.2.

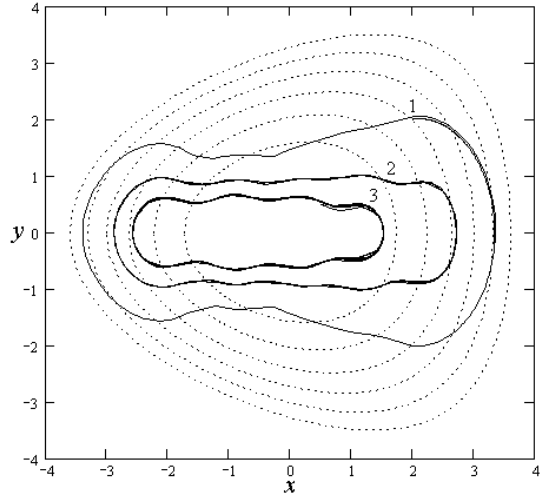

**Figure 4.** Trajectories of second type starting from the initial points: 1) x=2, y=2; 2) x=1, y=1; 3) x=0.5, y=0.5

The most important thing is that it is clearly seen now that trajectory increases its potential energy after launching (see Fig.3). Thus we proved our initial supposition about barrier tunneling in process of dynamical tunneling in coupled systems.

## IV. WATER POTENTIAL

In a study of vibrational states of the water molecule in the ground electronic state, where dynamical tunneling has been discovered, Lawton and Child employed the most realistic potential proposed by Sorbie and Murrel [10]. Many authors used another type of potential due to its simple analytical form but it is still able to reproduce a good agreement with experimental data up to five quanta of stretch excitation. For a fixed bond angle this potential is given [11]

$$U(x, y) = D(1 - \exp(-\alpha x))^2 + D(1 - \exp(-\alpha y))^2 + Fxy \quad (9)$$

The dimensionless parameters used here are $D=11.86$; $\alpha=0.205$; $F=-0.013$. Besides, the kinetic energy is not diagonal in this coordinate system, therefore the Hamiltonian is

$$\hat{H} = \frac{1}{2}\hat{P}_x^2 + \frac{1}{2}\hat{P}_y^2 + \mu\hat{P}_x\hat{P}_y \cos\Theta + U(x, y) \quad (10)$$

where $\mu = \frac{m_H}{m_H + m_O}$, $m_H$ and $m_O$ atomic mass of hydrogen and oxygen, respectively; $\Theta=104.52^o$ is a fixed bond angle. The eigenstates for the Hamiltonian (10) come in pairs as for previous case of coupled harmonic potential. The superposition of one of the pairs is shown in Fig.5.

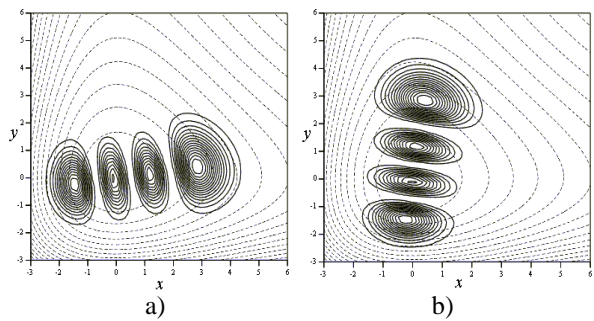

**Figure 5.** Superposition of the sixths and sevenths vibrational states ($E_6$=3.726, $E_7$=3.73) at t=0 (a) and t=$\tau$ (b).

It is harder to perform a trajectory calculation for this system than for coupled harmonic potential due to low energy splitting even for the lowest vibrational states. Consequently, the period of tunneling (8) is very high. So the system (4) can be integrated only for a short time.

Employing the similar calculation procedure, we obtained quantum trajectories presented in Fig.6. So one can conclude that the barrier tunneling is typical for this kind of potential, too.

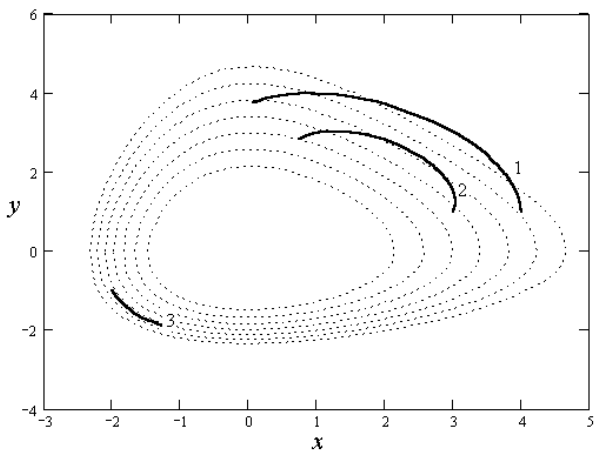

**Figure 6.** Quantum trajectories starting from the initial points: 1) $x(0)=1$, $y(0)=4$; 2) $x(0)=3$, $y(0)=1$; 3) $x(0)=-2$, $y(0)=-1$

Nevertheless, at some initial condition trajectory can follow the path where potential does not increase. At least, it is rightly within allowable integration time. Anyway, an existence of small portion of barrier tunneling trajectories means that energy splitting is due to this effect in dynamical tunneling process.

## V. SUMMARY

Tunneling had been associated for a long time with a barrier penetration. It was reconsidered since the discovery of dynamical tunneling. Heller proposed to define the tunneling as processes which take place in a quantum world but are forbidden in classical domain regardless of potential barrier availability.

In two-dimensional coupled potentials considered in current work, classical system can be localized in some spaces. Its dynamics prohibits it from making transition to other places though there is no energy restriction. Quantum system can make such a transition. As we have shown, it is not due to its motion along the energy valley on the energy surface. Instead, the system passes through the regions with higher potential energy. In other words, barrier penetration occurs during dynamical tunneling. It can explain specific energy splitting for these systems.

[1] R.T.Lawton and M.S.Child, Mol.Phys. **37**, 1799 (1979)
[2] R.T.Lawton and M.S.Child, Mol.Phys. **40**, 773 (1980)
[3] M.J.Davis and E.J.Heller, J.Chem.Phys. **75**, 246 (1981)
[4] E.J.Heller, J.Phys.Chem. **99**, 2625 (1995)
[5] E.J.Heller, J.Phys.Chem.A **103**, 10433 (1999)
[6] L.D.Landau and E.M.Lifshitz, *Quantum Mechanics. Non-relativistic Theory* (Pergamon Press, New York, 1977)
[7] D.Bohm, Phys.Rev. **85**, 166 (1952)
[8] D.Bohm, Phys.Rev. **85**, 180 (1952)
[9] P.R.Holand, *The Quantum Theory of Motion* (Cambridge University Press, Cambridge, 1993)
[10] K.S.Sorbie and J.N.Murrell, Mol.Phys. **29**, 1387 (1975)
[11] J.Zhang and D.G.Imre, J.Chem.Phys. **90**, 1666 (1989)